# Stakeholder-driven research in the European Climate and Energy Modelling Forum

Emir Fejzic[1,*], e-mail: fejzic@kth.se, ORCID iD: https://orcid.org/0000-0002-2489-8455
Will Usher[1], e-mail: wusher@kth.se, ORCID iD: https://orcid.org/0000-0001-9367-1791

[1] KTH, School of Industrial Engineering and Management (ITM), Energy Technology, Energy Systems, Brinellvägen 68 SE-100 44 Stockholm, Sweden
[*] Corresponding author details: E-mail address: fejzic@kth.se

## Abstract

A fast-paced policy context is characteristic of energy and climate research, which strives to develop solutions to wicked problems such as climate change. Funding agencies in the European Union recognize the importance of linking research and policy in climate and energy research. This calls for an increased understanding of how stakeholder engagement can effectively be used to co-design research questions that include stakeholders' concerns. This paper reviews the current literature on stakeholder engagement, from which we create a set of criteria. These are used to critically assess recent and relevant papers on stakeholder engagement in climate and energy projects. We obtained the papers from a scoping review of stakeholder engagement through workshops in EU climate and energy research. With insights from the literature and current EU climate and energy projects, we developed a workshop programme for stakeholder engagement. This programme was applied to the European Climate and Energy Modelling Forum project, aiming to co-design the most pressing and urgent research questions according to European stakeholders. The outcomes include 82 co-designed and ranked research questions for nine specific climate and energy research themes. Findings from the scoping review indicate that papers rarely define the term 'stakeholder'. Additionally, the concepts of co-creation, co-design, and co-production are used interchangeably and often without definition. We propose that workshop planners use stakeholder identification and selection methods from the broader stakeholder engagement literature.

**Highlights**

- We identify criteria for investigating the application of stakeholder engagement: stakeholder definition, identification, selection, workshop purpose and engagement type.
- There is a disparity between the profile that stakeholder engagement has in EU proposals, and the small number of papers identified in our scoping review that formally describe stakeholder engagement through workshops.
- The terms stakeholder, co-creation, co-design, and co-production are often not defined and sometimes used interchangeably in current EU climate and energy research
- Bias mitigation approaches in the identified literature are lacking despite work with high uncertainty
- We present a process for developing workshops within EU climate and energy modelling projects to co-design research questions.

# 1. Introduction

A fast-moving policy context characterises the energy and climate mitigation field, highlighting the need for public support for action. The policy context includes the impact of wars on supply chains of green technologies and bridge fuels such as natural gas [1] , surging energy prices, and energy security challenges [2]. While these challenges adversely affect the energy transition, changes in public support for clean energy and calls for fossil-fuel phase-outs have emerged due to the Russia-Ukraine war [3]. However, public support must be translated into policy action to affect the transition. In recent years, the policy landscape has changed to meet the existential threat of the climate crisis and the effects of the Russian invasion of Ukraine through the RePowerEU plan [4]. The changes include the introduction of the Inflation Reduction Act in the United States [5] and the European Green Deal in the European Union [6].

The Green Deal is the EUs core policy package to reach climate neutrality by 2050 [6]. The EU has aligned scientific programmes such as Horizon 2020 (European Commission, 2022) and Horizon Europe (European Commission, 2023a) with the Green Deal to help achieve its commitments. Creating solutions to prevent a climate crisis is a highly complex task. Climate change is commonly called a 'wicked problem' [7, 8]. 'Wicked problems' and 'post-normal science' require scientists to work closely with policymakers and citizens, given the high stakes involved and system complexity [9, 10].

In recognition of the importance of better linking research and policy in energy and climate, the European Climate and Energy Modelling Forum (ECEMF) project [11], funded through the Horizon 2020 programme aims to build a coherent evidence base to support policymaking. A key aspect of the ECEMF stakeholder engagement approach is co-designing research questions through workshops. This ensures relevance in the rapidly changing societal and technological landscape related to energy and climate research. We use the term co-design based on [12] defining it as "co-design describes active collaboration between stakeholders in designing solutions to a pre-specified problem".

As evidenced [13], funding agencies have long required projects to include stakeholder engagement activities at different stages. Additionally, stakeholders are essential in identifying socially relevant and scientifically challenging research questions. Therefore, they should be included in EU climate and energy research.

This study defines stakeholder engagement as an iterative process of dialogue and communication between stakeholders and organizations, aiming to achieve a specified objective. In our approach and the reviewed literature, we focus on studies using multi-stakeholder workshops as the engagement method. The general and well-established definition of the term stakeholder is "any group or individual who is affected by or can affect the achievement of an organization's objectives" [14]. We define stakeholders as a person or group with a direct interest or relation with the transition to climate neutrality in the European Union. It includes but is not limited to individuals or groups, such as national and international policymakers, researchers, private sector representatives or representatives of non-governmental institutions.  As we report in the scoping review in Section 3, there is little agreement about who is considered a stakeholder in EU climate and energy research.

This article presents a scoping review of stakeholder engagement in EU climate and energy research. We critically assess the implementation of stakeholder engagement, as outlined by the existing methodological literature, scrutinizing how these methodologies have been employed in recent climate and energy projects. We present and appraise a methodology for stakeholder engagement drawing on experiences from the ECEMF project and the available literature. The research approach of this paper is both applied and exploratory. Our strategy does not follow a pre-defined framework or methodology. Still, it takes bits and pieces from other research papers and applies relevant parts to our case study, i.e. the ECEMF workshop program.

While stakeholder engagement using workshops is well established in climate and energy research, there is no agreed handbook for multi-stakeholder workshops in European energy and climate projects. Although many projects purport to incorporate some form of stakeholder engagement, there is little critical reflection on the nature of this engagement. There are few reviews of the state-of-the-art and limited specific guidance for European climate and energy research regarding stakeholder engagement. Anecdotally, there seem to be differing standards and expectations across disciplines, from informal gatherings to very formal, structured affairs with strictly controlled participation.

In this article, we aim to bring clarity to the above-stated shortcomings, providing answers to the following two research questions:

- How has stakeholder engagement been conducted in the European climate and energy space?
- What attributes, approaches and techniques could make stakeholder engagement more effective?

In Section 2, we investigate the theoretical background of stakeholder engagement. In Section 3, we review recent European Union-funded energy and climate projects to understand current practices. In Section 4, we describe the ongoing ECEMF approach that has been conducted in parallel to this study using an iterative and reflective learning-by-doing process. In Section 5, we discuss the results from these workshops and reflect on our lessons for future European-funded energy and climate projects that plan to implement stakeholder engagement activities.

## 2. Literature Review

In this section, we investigate what stakeholder engagement is, touch on the importance of connecting scientific knowledge on energy and climate with actors outside of science and review the stakeholder identification and selection process from the literature. We conclude with a list of criteria drawn from this review that we subsequently use to assess existing approaches to stakeholder engagement in European climate and energy projects (Section 3).

Stakeholder engagement, also referred to as stakeholder participation, has undergone developments in parallel geographical and disciplinary contexts since the 1960s (for a comprehensive review, see [15]). This has led to confusion regarding what stakeholder engagement entails, who is considered a stakeholder, and what steps should be taken for effective engagement [16]. Stakeholder participation has been described by [17] as "*used by a variety of individuals, supported by a range of contradicting evidence and arguments*". These claims are supported by recent publications in the field of stakeholder engagement, with [18] highlighting that the stakeholder engagement construct is a broad term with many, sometimes diverging, definitions.

We see stakeholder engagement as an iterative process that entails ongoing dialogue and communication between stakeholders and organizations to achieve a specific, pre-defined objective. It requires the organization to be responsive to stakeholder needs and to seek active feedback from stakeholders. The idea that stakeholder engagement should be a continuous, iterative process between the organization and the stakeholders is supported by [19] and [20], who calls for the involvement of stakeholders at all project cycle stages.

Benefits of stakeholder engagement for knowledge co-production include different perspectives, interpretation of findings, and scrutiny of assumptions [21, 22] cited in [23]. The reasons for engagement include an increased sense of ownership, transparency and trust, and incorporation of stakeholder knowledge, to name a few [24]. Additionally, co-creation practices are advocated for by [25] for managing complex systems when uncertainty is present. Climate change is commonly referred to as a 'wicked problem' [7, 8]. 'Wicked problems' and 'post-normal science' require scientists to work closely with policymakers and citizens, given the high stakes involved and system complexity [9, 10]. Knowledge production in the global environmental change context should be based on collective problem framing and a wide range of perspectives [26]. A separate push for further engagement with scientific knowledge arises from the principles of Open Science, which recognise the value of working with "*societal actors beyond the scientific community*" for "*enhanced dialogue between scientists, policymakers and practitioners, entrepreneurs and community members*" [27].

As mentioned by [28], policymakers influence models and modellers, and engaging policymakers and stakeholders increases the model's impact on policy output. This is the preferred outcome and of the engagement process within the ECEMF. Stakeholder involvement in the co-design process of establishing relevant and urgent research questions motivates the stakeholder engagement process.

Co-creation, co-design, and co-production are common concepts associated with stakeholder engagement activities [29-32], often described differently while sometimes treated synonymously.

Examples include differing definitions of co-production [31, 33, 34], and empirical evidence suggests that co-creation and co-production often appear interchangeable in public innovation [35], with little focus on the outcomes of the process. Work published by [36] highlighted this phenomenon almost two decades ago, and more recent work confirmed that this remains the case [37]. Co-creation and co-production are required to develop a sense of ownership among stakeholders regarding policy narratives that transform climate action [38]. Previous studies also show the use of knowledge co-production through stakeholder workshops to extract stakeholder knowledge [39] and to create decarbonization pathways [40].

Few attempts have been undertaken to establish clear differences and applications between these concepts [12]. Our work refers to the process of active collaboration, undertaken at each workshop, as *co-design*. The choice is based on the definition provided by [12] stating that "*co-design describes active collaboration between stakeholders in designing solutions to a pre-specified problem*". The approach also aligns with the reasoning that co-design and co-production fall under the broader term of co-creation, with each concept having its own distinct aspects but also shared commonalities [12, 36]. Furthermore, studies have demonstrated the importance of collaborating with stakeholders early in the co-creation process, through co-design [41], which aligns with the timing of our co-design implementation.

Stakeholder identification is the first step of stakeholder engagement, as shown in the different frameworks and approaches presented by [16, 42-44]. Sometimes, the starting point of stakeholder engagement is defined as a consultation step for co-defining or co-designing pertinent topics or alternative narratives for energy system pathways [45, 46]. Identifying stakeholders requires a clear definition of the term stakeholder. Many definitions exist, e.g. [47-49], however, there is no consensus within the field of energy and climate research on what constitutes a stakeholder. In this work, we define stakeholders as a person or group with a direct interest or relation to the transition to climate neutrality in the European Union.

The second step includes differentiating and categorizing stakeholders, as presented by [16]. Categorization can be performed using analytical categorization (top-down) or reconstructive categorization. Examples of analytical categorization include stakeholder classification based on their knowledge and function [43], by interest-influence matrices [42, 44, 50], or through the creation and inclusion of functional criteria and stakeholder roles combined with the dimensions of interest and influence [23, 51]. The reconstructive categorization method stems from reconstructive democratic theory and allows stakeholders to express their interactive competencies and categorize themselves, thus giving analysts less interpretive latitude [52]. It builds on the categorization undertaken by stakeholders as opposed to the analytical categorization normally developed by scientists. This approach is also known as stakeholder-driven stakeholder categorization [53].

When selecting stakeholders or conducting any sampling of data smaller than the total population of data points/people, one must apply some sampling strategy. When maximised sample variation is desired, stakeholders can be matched with other stakeholders as different to them as possible. The differences can include age, gender, stage in their careers, geographic location, ethnicity, and nationality. Selecting a sample that contains maximised sample variation provides higher-quality data and insights into important stakeholder commonalities [54]. Selection methods have been developed to support organizers of stakeholder engagement in this process, including [55] cited in [56]. Both [54] and [55] emphasize the challenge of representativeness, highlighting the importance of stakeholder identification and selection in public engagement processes.

From reviewing the literature, we derived a list of criteria, based on key concepts that emerge from the review. We used them to critically assess existing approaches to stakeholder engagement within European climate and energy projects in section 3. With these criteria in mind, we present the approach to stakeholder engagement in the ECEMF project in section 4. This provides novel perspectives on effectively engaging stakeholders in co-designing research questions within climate and energy research.

- How do workshop organisers define stakeholders?
- What stakeholder identification method do they use?
- What stakeholder selection method do they apply?
- How do the organisers define the purpose of the workshop? Co-creation/co-development/Co-design/engagement/dissemination of results/other/not stated?
- How does engagement take place in workshop settings?

# 3. Scoping Review: Stakeholder Engagement in EU Climate and Energy Projects

This study adopted a cross-sectoral research design to study stakeholder engagement in EU climate and energy projects. The paper draws on a scoping review of recently published work from 2018-2023 and 3 workshops on energy and climate in the EU context conducted between 2021 and 2023 as part of the ECEMF project.

## 3.1. Scoping review

Various methods exist for engaging stakeholders and policymakers in co-designing research questions and disseminating results of research projects. We perform a scoping review [57] to better understand how stakeholder engagement is achieved through workshops in EU climate and energy projects and to identify the knowledge gaps. The approach draws inspiration from previous developments in systematic scoping reviews [58]. The review includes (1) a definition of the review questions, (2) a definition of exclusion criteria, (3) the creation of a search string, (4) article screening, and (5) data extraction. The review questions include: (1) What methods have been implemented for stakeholder engagement in European climate and energy projects? (2) What recommendations exist on how to conduct stakeholder engagement in European climate and energy projects? (3) What level of implementation of stakeholder engagement exists in European climate and energy-related projects?

The second step includes the definition of exclusion criteria. These are used categorically to exclude articles outside the review scope. The following three types of exclusion criteria are used:

- Geographical scope: Studies with a geographical scope other than the EU.
- Sector/research field of analysis: Studies covering sectors or research fields unrelated to energy and climate.
- Stakeholder engagement approach: Studies with primary stakeholder engagement methods other than workshops.

With the review questions and exclusion criteria established, we created the search string, shown in Table 1. The selected language is English, and we focused on literature published between 2018 and 2023. The review was limited to open access papers as the European Commission supports open access, particularly as part of its funding programmes. The EU's open science policy also aims to make open data sharing the norm for EU-funded research and open access to research data and publications [59]. The search was conducted in June 2023.

Table 1 Scoping review search string.

| Platform | Database | Search string |
|---|---|---|
| Web of Science | Core Collection | ALL=((Workshop OR stakeholder) AND (EU OR European union OR European OR Europe) AND ENERGY AND (INDUSTRY OR RESEARCH OR TRANSPORTATION OR MARKETS)) |

The Web of Science (WoS) search resulted in 1463 articles. One paper was manually added. The journal in which it was published was not indexed by the WoS, and did, therefore, not show up in the search. After removing duplicates, the number of unique articles was 1463. The screening process was divided into two stages: (1) checking the eligibility of articles based on title and abstract and (2) checking the eligibility based on full-text reads. After the first screening stage, the number of selected articles was 76, while 1387 were rejected based on the exclusion criteria. At the full-text retrieval stage, we identified three papers we could not access. These were excluded, leaving 73 papers retrieved in full text.

The final step of the scoping review was to check eligibility based on full-text reads. Results from the scoping review are presented in section 3.2.

### 3.2. Scoping review findings

From the 73 articles retrieved in full text, we identified 11 as relevant, while 62 were excluded. The included articles cover the topic of stakeholder engagement using workshops in EU climate and energy research. The scoping review process is shown in Figure 1 as a PRISMA (the Preferred Reporting Items for Systematic reviews and Meta-Analyses, PRISMA) flow chart.

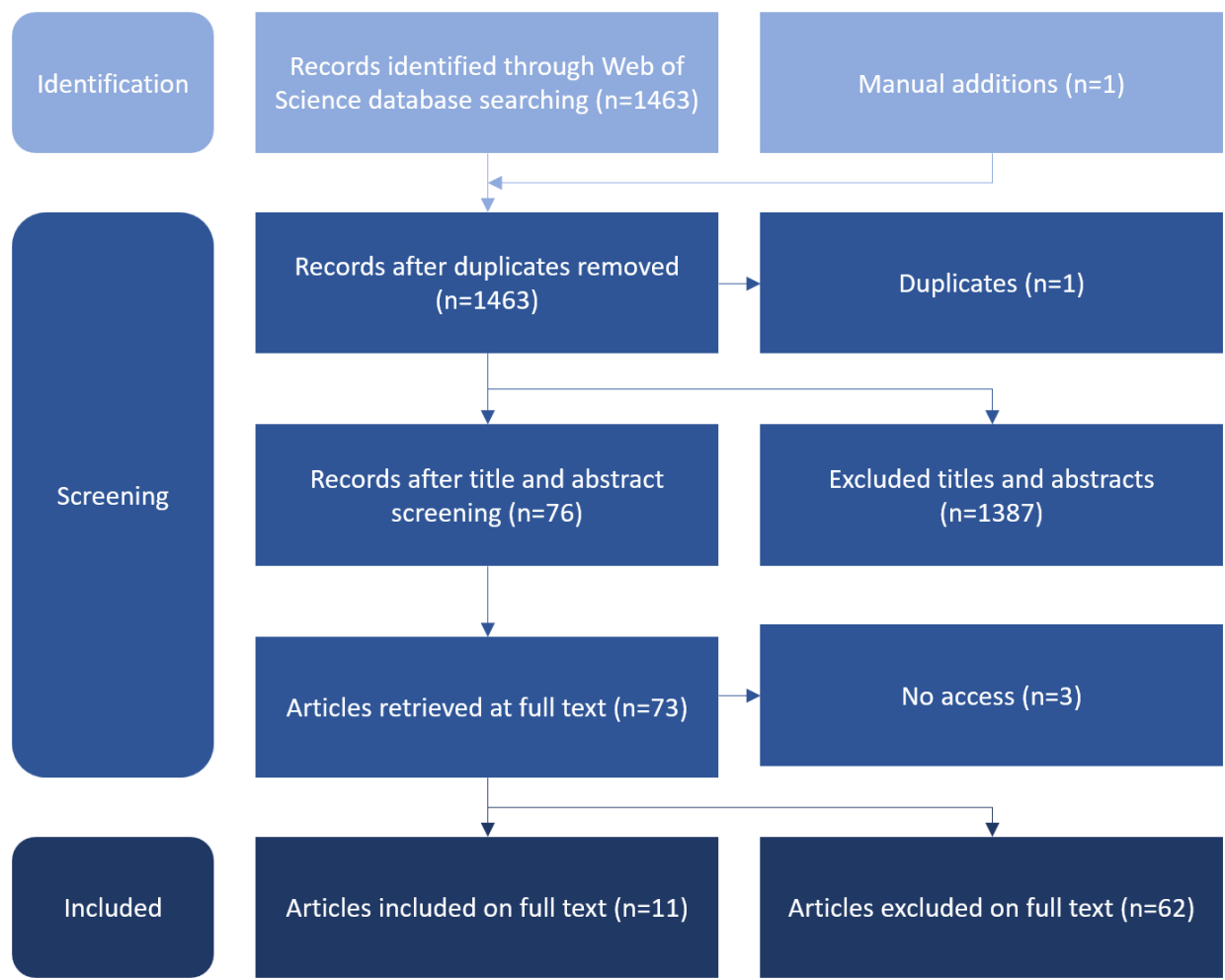

Figure 1. PRISMA flow chart illustrating Web of Science search results and articles excluded or included during the review stages.

We used publications from ongoing and recently completed EU-funded projects to validate the selection of included scoping review articles. We examined the publications of six selected EU-funded projects related to energy or climate. These projects include CINTRAN[1], EnerMaps[2], NAVIGATE[3], ParisReinforce[4], SENTINEL[5], and WHY[6]. Among 204 articles published in these six projects, 17 cover stakeholder engagement, four of which address stakeholder engagement through workshops, emphasising energy or climate within the EU. The findings of the scoping review were deemed robust as all four articles identified in the six selected projects also appeared included in the scoping review.

---

[1] CINTRAN: doi.org/10.3030/884539
[2] EnerMaps: doi.org/10.3030/884161
[3] NAVIGATE: doi.org/10.3030/821124
[4] ParisReinforce: doi.org/10.3030/820846
[5] SENTINEL: doi.org/10.3030/837089
[6] WHY: doi.org/10.3030/891943

A summary of the stakeholder engagement achieved through workshops in EU climate and energy research is presented in Table 2. It describes the similarities and differences between the approaches and scope of the 11 full-text articles included.

Table 2. Summary of stakeholder engagement through workshops in EU climate and energy projects.

| Article | Main engagement method(s) | Geographical scope | Sector/ Research area |
|---|---|---|---|
| [40] | • 5 online workshops | • EU<br>• Global<br>• Asia<br>• Latin America | Decarbonisation pathways |
| [60] | • 53 interviews<br>• 7 workshops | • EU and local levels | Energy service business models (ESBMs) |
| [61] | • Online survey<br>• Meeting, interviews & focus groups.<br>• 2 Online workshops | • EU<br>• Regional<br>• National | Energy demand modelling |
| [62] | • Online survey<br>• 1 meeting including interviews & focus groups.<br>• 2 Online workshops | • EU<br>• Regional<br>• National | Energy demand modelling |
| [63] | • Interviews<br>• Workshops<br>• Systematic horizon scanning | • EU | Social sciences and humanities in renewable energy sources research |
| [45] | • 2 workshops | • EU | Energy modelling |
| [64] | • 32 interviews<br>• 1 survey<br>• 1 workshop | • EU | Energy modelling |
| [46] | • 1 workshop | • EU | Energy modelling |
| [65] | • 1 survey<br>• 2 workshops | • EU | Social Sciences and Humanities |
| [56] | • 2 surveys<br>• 2 workshops | • EU | Methodology development for climate action |
| [66] | • 5 in-person workshops | • EU | European Renewable energy sector |

As mentioned in section 2, we apply a set of criteria to access critically the articles shown in Table 2. In Table 3, we present a summary of the findings. Brief answers are provided to the criteria questions to evaluate the implementation of theoretical stakeholder engagement concepts in EU climate and energy projects.

Table 3. Assessment of included scoping review articles.

| Author | Definition of the workshop purpose | Engagement approach |
|---|---|---|
| [40] | Co-production of project results, co-design, and co-production of decarbonization pathways, dialogues on national policies and pathways. | Presentations, posters, group discussions, panel discussions, surveys. Use of an interactive visual tool |
| [60] | The workshops were conducted to explore contemporary domestic ESBM examples in Europe. | Interviews; Workshops; Business Model Co-production Method; Business Model Canvas |
| [61] | To discuss, verify, and prioritize a preliminary set of general user needs identified through an online survey. | Online survey, stakeholder meetings and workshops. |
| [62] | To identify critical issues and challenges of the EU energy transition as perceived by stakeholders | Online workshop and online interviews; focus groups |
| [63] | Utilising expert knowledge of researchers from the field with diverse disciplinary backgrounds and expertise. | Interviews to find working group members; Survey to collect research questions from researchers; Working group deliberate workshops for consideration of top-ranked research questions |
| [45] | The use of workshops aimed to explore how large-scale trends, technology pathways, and policy options interact to shape Europe's energy future through co-designing. Co-design and co-development used interchangeably. | Survey, two workshops including stakeholders from the project and identified experts. |
| [64] | To identify and prioritize user needs for energy models. | Online interviews to extract sets of relevant environmental and social aspects; Survey; Workshop including breakout sessions and live polls. |
| [46] | Co-designing scenarios and research questions, co-production of research. Terms co-design, co-create, co-production and co-defining are used without any definition to these concepts. | Presentations and discussions concerning a policy brief; Workshop based on thematic topics; Voting process via sli.do, including prioritization of research questions. |
| [65] | Engaging stakeholders for the purpose of reviewing, commenting on, and discussing a set of documents. | Participants took part in two workshop plenaries. |
| [56] | Co-creation of pathways for European sustainability. No definition is provided for the concept of co-creation. | Surveys; Two workshops for discussing and creating stakeholder strategies through moderated backcasting. |
| [66] | To co-produce and co-design an online interactive tool, with external and internal stakeholders. The term co-production and co-design are used interchangeably. | Workshop including participatory exercises - User Stories; hands-on sessions building data assessments; presentations; interactive sessions using the tool; Webinars; Advisory committee |

| Author | Definition of the term stakeholder | Stakeholder identification approach | Stakeholder selection approach |
| --- | --- | --- | --- |
| [40] | Term 'stakeholder' not defined. Stakeholder categorization of invited stakeholders applied. | Follows the (Gramberger et al., 2015) Prospex-CQI method for stakeholder identification and selection. | Prospex-CQI by (Gramberger et al., 2015) |
| [60] | Term 'stakeholder' not defined. Stakeholder categorization of invited stakeholders applied. | Identification approach not specified | Not specified |
| [61] | Term 'stakeholder' not defined. Stakeholder categorization of invited stakeholders applied. | Identification approach not specified | Not specified |
| [62] | Term 'stakeholder' not defined. | Identification approach not specified | Not specified |
| [63] | Term 'stakeholder' not defined. | Identification approach not specified | Not specified |
| [45] | "A person or group that has a direct interest or relation with the decarbonization process. It embodies actors from a variety of fields, including research, business, finance, industry associations, academia, national and EU policy makers." | Identified through networks of the project participants, contacts with government agencies, and online searches. | Stakeholders selected based on their experience |
| [64] | Term 'stakeholder' not defined. However, stakeholder categorization applied for the conducted interviews. | Stakeholders identified through a document analysis based on policy documents and government-commissioned studies. | Not specified |
| [46] | Term 'stakeholder' not defined. | Identification approach not specified | Stakeholders invited based on displayed participation or interest in European climate-policy events. |
| [65] | Term 'stakeholder' not defined. However, the paper uses the term "human actors", which encompasses stakeholders among others. | Identification approach not specified | Not specified |
| [56] | Term 'stakeholder' not defined. | Follows the (Gramberger et al., 2015) Prospex-CQI method. | Prospex-CQI by (Gramberger et al., 2015) |
| [66] | 'Stakeholders' referred to as ECEM project team and external stakeholders. The external stakeholders defined as users and potential users of the Demonstrator and other ECEM outputs. | Stakeholders identified based on existing networks of team members. | Not specified |

The review results indicate that many articles cover stakeholder engagement within energy and climate research within the EU, however, the literature regarding workshops as the primary method of engagement is limited. Only 11 out of 1,463 articles identified fit the review's scope and inclusion criteria. The included articles present workshops as the primary engagement approach, with a clear impact of the COVID-19 pandemic on how these workshops are organised. From the selected articles, six present in-person workshops, while five follow an online workshop setting. We observe that articles published after 2020 [45, 46, 60, 65] build on results from workshops conducted in 2018 and 2019. All in-person workshops were conducted before the pandemic. Workshops presented in articles published after the pandemic were held online. This confirms [67] findings highlighting a shift from in-person workshops to an online format as a consequence of the pandemic. The review highlighted limited scientific literature available on stakeholder engagement through workshops in EU climate and energy research.

Table 3 shows that only two articles state a specific definition of the term 'stakeholder'. Four articles present stakeholder categories by which they categorize the attendees of the engagement activities. Articles where authors make specified claims as to what constitutes a 'stakeholder' provide varying definitions. The results confirm the findings presented in section 2, showing that there is no consensus on what constitutes a stakeholder. Based on this criterion, we conclude that this holds true for European climate and energy projects as well.

Considering stakeholder identification, the review results indicate that six articles do not specify any stakeholder identification approach. Three articles highlight their approaches for stakeholder identification, none of which follows a specific methodology designed for stakeholder identification. Finally, two articles follow the Prospex-CQI method [55]. The review highlights that in climate and energy research, stakeholder identification approaches are fragmented, with few studies implementing strategies based on methods described in stakeholder engagement literature.

As projects are unique and have different needs, the stakeholder identification process often requires modifications. Utilizing methods such as Prospex-CQI, may help reduce bias. In contrast, developing research questions and identifying pressing research needs, solely relying on existing networks of team members, may lead to an increased risk of representation bias.

In addition to identification, stakeholder selection plays a crucial role in ensuring a diverse representation of stakeholders. The review results indicate that only four articles present a selection approach, two of which follow the Prospex-CQI method, while two focus on the stakeholders' interests and experiences. While stakeholder interests are part of the *interest-influence matrix [15]*, we observe no applications of methods such as radical transactiveness [68], stakeholder-led stakeholder categorisation [53], or Q methodology [69] among identified articles. These methods for differentiating and categorizing stakeholders are well established in the stakeholder engagement literature but appear not to be utilized in climate and energy research.

Approximately half of the identified articles used co-creation, co-production, co-design, or a combination of two or more of these concepts as the main reasons for the stakeholder engagement.

The concepts are used interchangeably and without a clear definition as to what they entail. This confirms the findings from the stakeholder literature presented in section 2.

All identified articles present interactions between internal project members and external stakeholders. They use differing approaches to communicate their needs, ideas, and opinions on pressing issues. Although the articles deal with wicked problems and work with high uncertainty, no bias mitigation approaches are mentioned.

Online workshops allow the use of tools such as Zoom, where participants can be divided into different virtual rooms based on their expertise, topics of interest, or other differentiating factors. Three articles included in the scoping review share common workshop events. However, they address different aspects and draw inputs from diverse breakout sessions. While [61, 64] share the workshop for user needs in energy modelling, [61, 62] share the workshop when regarding case studies on regional and continental levels. Resource sharing and collaborating on workshop activities while extracting information during breakout sessions could prove helpful in climate and energy projects that cover a broad range of topics.

# 4. The ECEMF approach to co-design of research questions

In this section, we present the design and application of the stakeholder engagement approach taken in the first three workshops of the ECEMF project, including the outcomes and main findings of the co-design process.

## 4.1. Overview of the ECEMF Workshop Process

The ECEMF objectives include identifying and co-designing policy-relevant research questions, involving stakeholder engagement through active dialogue. It represents the first interaction between the ECEMF and external stakeholders. In Figure 2, these engagement activities are indicated in red. The methodology presented in section 4 connects to the project at this stage, focusing on stakeholder engagement to co-design policy-relevant research questions. Consequently, these interactions contribute to a comprehensive evidence base for assessing energy and climate policies, supporting the development of policy-relevant insights. Figure 2 shows this work in blue and constitutes internal project activities. It includes a model intercomparison exercise using a cohort of energy and integrated assessment models to answer the identified research questions. The second interaction phase, planned for upcoming workshops, includes communication with decision-makers through dissemination activities. The ECEMF project activities will ensure policy-relevant research, a comprehensive and coherent evidence base leading to robust policy insights, and the advancement of state-of-the-art energy and climate modelling. In Figure 2, the yellow section represents dissemination activities.

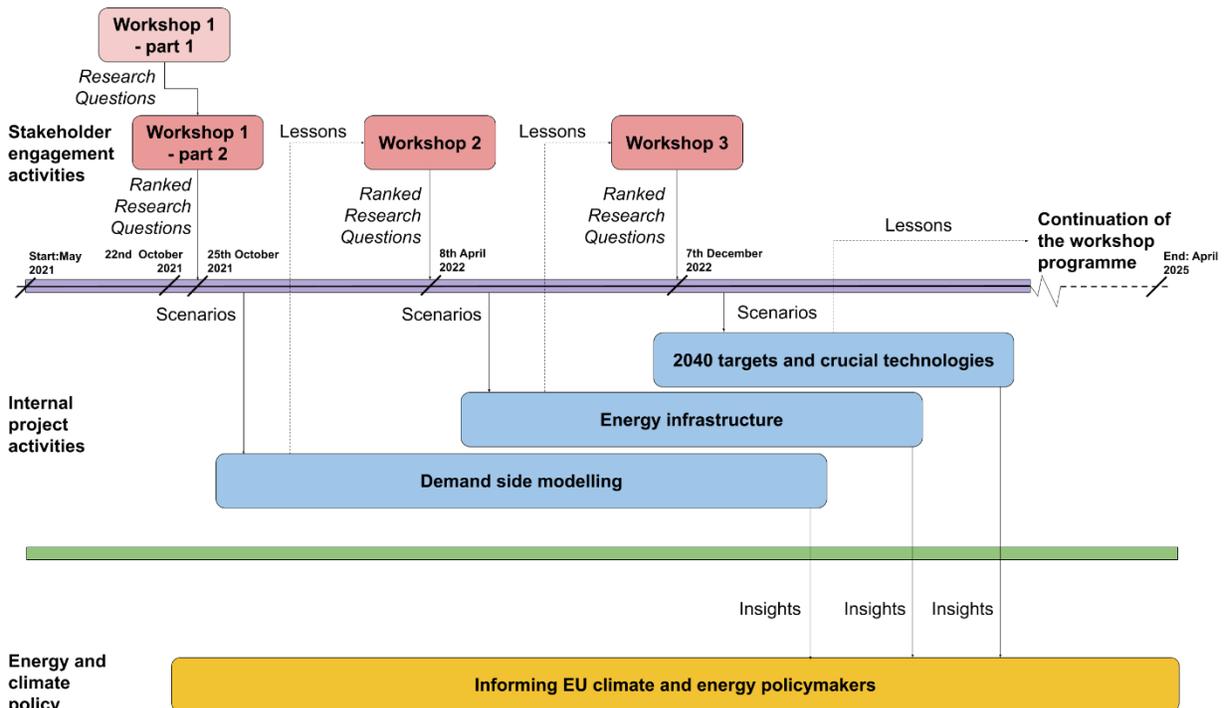

Figure 2. Diagram showing the relation between workshop phases, project research, and iterative learning feeding-forward.

The workshop program is set to consist of eight workshops between 2021 and 2025. Figure 2 illustrates the timeline of the workshops that have been conducted at the time of writing and how they connect to internal project activities. The workshop outputs are used for modelling efforts within the project, resulting in insights and recommendations to EU policymakers.

The first workshop was organised on the 22nd and 25th of October 2021. The topic covered energy demand and demand side narratives. The workshop had (n=15) participants on the first day and (n=14) on the second. The second workshop focused on energy infrastructure, and it took place on the 8th of April 2022, with (n=29) participants. The third workshop was held on the 7th of December 2022, with (n=27) participants attending. In total, 85 participants attended. The workshops were conducted online since they took place when the COVID-19 pandemic affected travel. As reported [67], this was a common practice among workshop organisers at that time.

The development the workshops included: stakeholder identification and selection, workshop implementation to co-design research questions, scenario creation and integration of questions into research, feedback, and dissemination. Sections 4.1.1 to 4.1.4 describe these steps in detail.

The overall workshop programme approach has developed through observation of issues and challenges during the workshops. Section 4.1.5 provides information on the integration of lessons into the design of upcoming workshops.

### 4.1.1. Stakeholder identification

We consider stakeholder identification as the first step in the engagement process following the methodology presented by [16]. The identification method used comprised of two parts. Firstly, online questionnaires were used to identify relevant stakeholders for each workshop. The questionnaires were created using Mailchimp and Google Forms and distributed through the LinkedIn and X platforms. Recorded number of interactions with the content on LinkedIn was 4,000. Quantifying interactions on X was not possible. This approach enabled reaching a large target audience while making use of project member networks. Secondly, we applied a chain referral technique called "Snowball sampling" [70, 71]. A vital component of this approach was internal brainstorming activities with project partners who provided referrals to stakeholders. We created list of identified stakeholders and invitees were requested to refer others interested in attending the workshops.

Although effective for starting a chain-referral process, brainstorming events with project partners may lead to representative bias [16]. This can be mitigated by applying stakeholder selection methods with pre-defined criteria to ensure diverse representation. Another approach is combining internal project brainstorming with online forms or questionnaires to reach a broader initial group of stakeholders. In our case, we applied the latter.

Depending on the workshop the stakeholders signed up for, they were asked to provide different types of information. In addition to personal information and contact details, they provided information about their institutional affiliation, sectoral background (e.g., oil & gas, renewable energy), and sector category (e.g., academia, private sector). The information was requested to support the stakeholder categorization and selection process.

### 4.1.2. Stakeholder selection

While stakeholder selection methods exist, e.g. [55] cited in [56] we did not apply any methods to select the workshop participants. The number of stakeholders who signed up for the workshops did not exceed what was considered manageable. Therefore, we included all stakeholders who signed up for the workshops. This included all respondents to the online questionnaires and forms shared before each workshop and people who expressed interest in participating via email as part of the snowball sampling. Our selection and categorization included dividing stakeholders into different breakout sessions during the workshops based on their backgrounds and expressed interests.

In total, 191 stakeholders were invited to attend the workshops, 85 of whom attended. The workshops were organised online using the Zoom platform, allowing people to attend regardless of geographical location and making the workshops more accessible. Reduced travel needs contributed to lower $CO_2$ emissions, as pointed out by [40]. The COVID-19 restrictions also resulted in positive trade-offs. The workshop organisers provided participants with Zoom links, enabling attendance without costs associated with software for online meetings. The sectoral background of stakeholders participating in the workshops is shown in Table 4.

Table 4. Sectoral division of the workshop attendees.

| Workshop | Academia | Research Institute | Intergovernmental / International organisation | Private Sector | Non-profit organisation | NGO |
|---|---|---|---|---|---|---|
| First | 12 | 8 | 8 | 1 | 1 | 0 |
| Second | 15 | 5 | 5 | 2 | 1 | 0 |
| Third | 9 | 7 | 5 | 5 | 0 | 1 |
| **Total** | **36** | **20** | **18** | **8** | **2** | **1** |

Achieving a comprehensive participant representation is crucial to ensuring diverse views on the challenges ahead. The gender balance of attendees was therefore monitored, with 64 men and 20 women attending the workshops. The gender distribution differed between the events, with most women attending the first workshop, corresponding to 40 per cent of the total attendees. The lowest share of female attendees was in the second workshop, with just 11 per cent of women participants due to last-minute cancellations attributed to other commitments by the stakeholders.

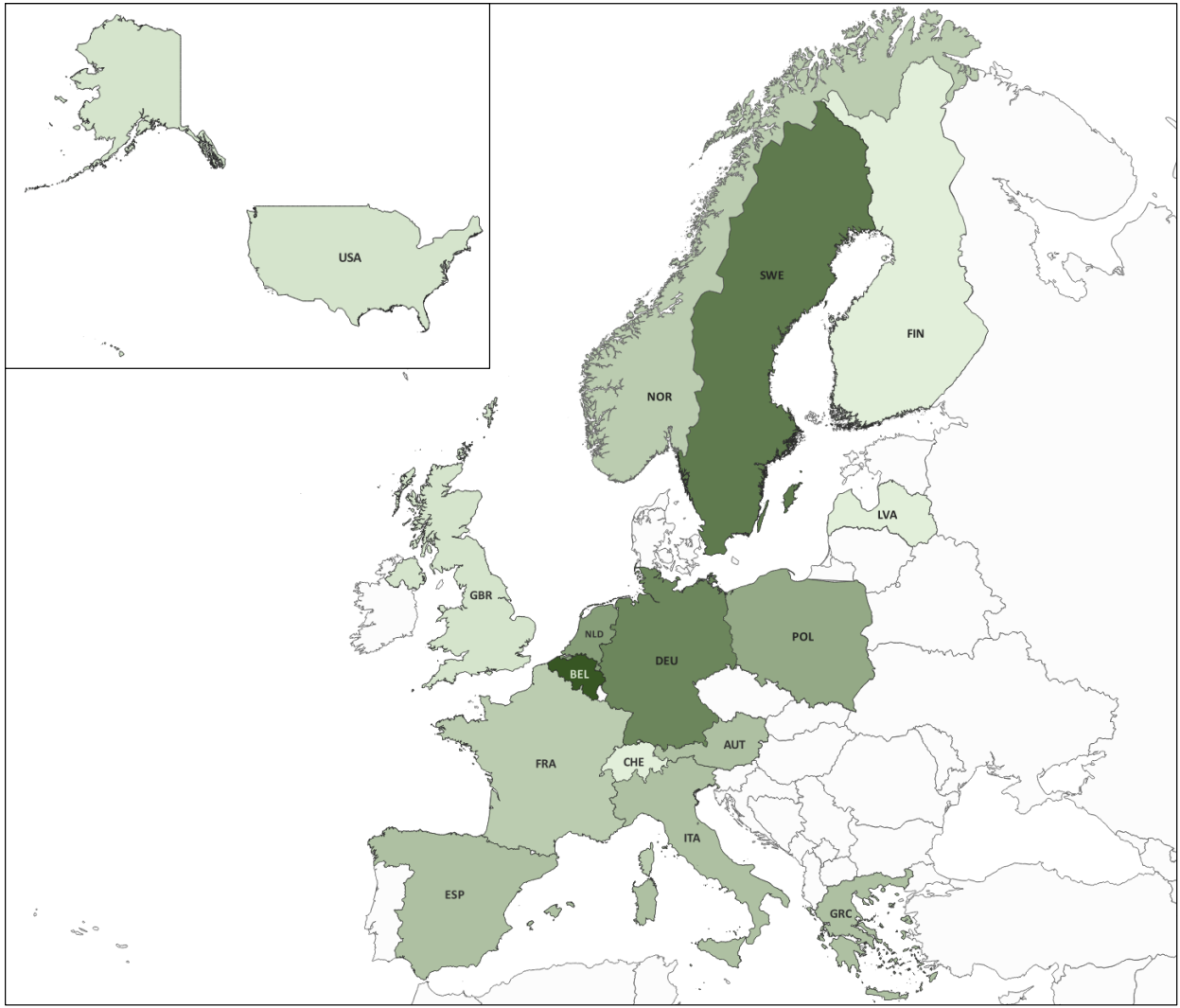

Figure 3. Country of affiliation of workshop participants. Darker colours indicate a greater number of participants.

### 4.1.3. Co-design of research questions

The workshops provided a meeting place for stakeholders and ECEMF representatives to co-design policy-relevant research questions for a range of thematic areas within climate and energy research. As some participants were high-level stakeholders from the European Commission, it was necessary to consider the timeframe of the workshops. Given stakeholders' busy schedules, assigning a long workshop format covering a full-day event could lower participation. Conversely, brief workshops create packed schedules with limited discussion time, potentially undermining intended aims. Through internal consultations, we presented a 2.5-hour format for the first workshop. The timeline was a limiting factor, providing insufficient time for the open plenary discussion. We then incorporated lessons learned from internal and stakeholder feedback loops to design a 3-hour workshop timeline, shown in Table 5, for the second and third workshops. These adjustments proved sufficient, as neither the invited stakeholders nor the ECEMF participants provided feedback on the format being too short.

Table 5. Timeline for a 3-hour workshop session applied in workshop 2.

| Activity | Allotted Time (minutes) | Clock Time |
|---|---|---|
| Introduction (facilitator introduction, agenda, tools used, etc.) | 10 | 0:00 |
| Presentation of key activities within ECEMF | 20 | 0:10 |
| Presentations by ECEMF task leaders for each research theme | 20 | 0:30 |
| Interventions (invited speakers) | 15 | 0:50 |
| Break | 5 | 1:05 |
| Breakout session (co-design of research questions) | 70 | 1:10 |
| Break | 5 | 2:20 |
| Open plenary discussion | 30 | 2:25 |
| Summary and closing remarks | 5 | 2:55 |

The workshop facilitator introduced the goals and objectives of the workshops, including information about the agenda and engagement tools needed. The tools included Mentimeter, which gathered stakeholder insights through a series of questions. This step served a two-fold purpose. Firstly, it provided the facilitators with additional information about the stakeholders. Secondly, it ensured that the stakeholders felt comfortable using Mentimeter and increased the sense of engagement early on. Stakeholder responses to a Menti icebreaker question are illustrated in Figure 4. The second tool was Miro, a digital collaboration platform used to engage the participants in the discussions and share ideas to co-design research questions. Both Mentimeter and Miro are accessible through a web browser, a mobile device, or a personal computer without the need for software installations.

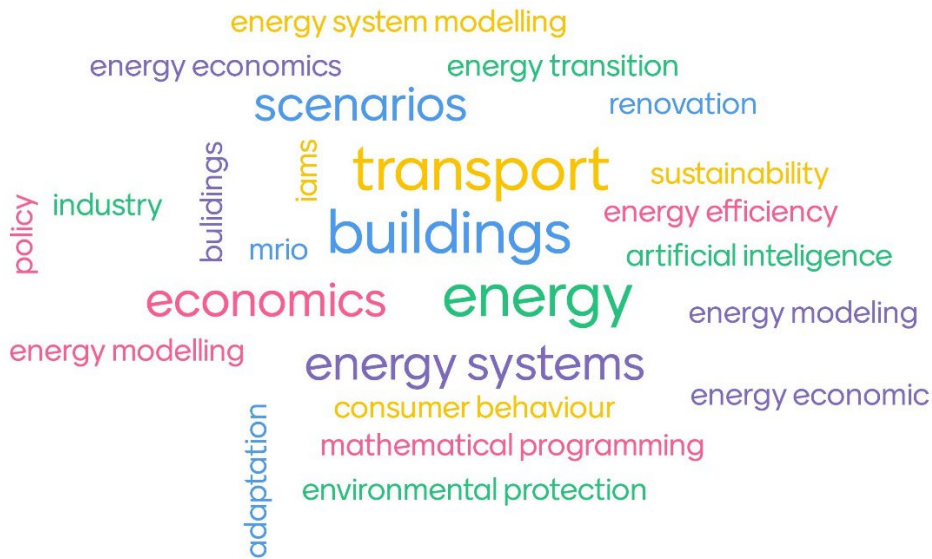

Figure 4. Word cloud of invited stakeholders' fields of expertise, workshop 1.

After presenting the workshop's objectives and using Mentimeter as an icebreaker, ECEMF members presented the key project activities. Task leaders for the covered research themes shared insights into the status of their work, including detailed information about models used in the ECEMF. This step was included to better inform stakeholders of the project scope and objectives.

Interventions from selected stakeholders were part of the agenda. This was to increase engagement and gain a better understanding of stakeholders' backgrounds and work. Stakeholders shared own experiences connected to ongoing ECEMF project activities and displayed their current work.

We divided the Zoom participants, allowing facilitators to provide focused attention to stakeholder groups whose interests aligned with the topics of the breakout sessions. Stakeholders ranked the questions using pre-constructed Miro boards to record their thoughts and comments. Using Miro in tandem with Zoom enabled participants to communicate via Zoom while simultaneously collaborating on designated Miro boards. The ranking process was based on policy relevance, urgency, and research significance. An example of the process is illustrated in Figure 5. A concluding session allowed stakeholders to share their views on the workshop content. It allowed for joint discussions of results.

| Research Questions | Policy relevance: Is the question policy relevant? | Suggested change: Suggested changes to make the question more policy relevant | Urgency: Which questions are the most urgent to answer? | Significance of research outcomes: What will we get out of it? | Total: Average of the post-it values |
|---|---|---|---|---|---|
| What is the role and contribution of energy efficiency improvement possibilities of internal combustion vehicles as an alternative? | 4 (PS); 4 (IT); 3 (P); 2 (SAS) | Negative: Prolongation of ICE technology; there are synergies with e-liquids; only if e-fuelled and more for road freight | 2 (IT); 2 (SAS); 3 (PS) | 2 (IT); 2 (SAS); 3 (PS) | 2.7 |
| What is the maximum potential for e-fuels in aviation and maritime? | 7; 7 (PS); 6 (P) | Suggestion: sector coupling (more clear about energy linkage) | 7; 3 (SAS) | 8; 8 (PS); 5 (SAS) | 6.375 |
| Under which conditions could we achieve a high modal switch from aviation to High speed rail? | 2 (P); 8 (IT); 8 (SAS); 8 (PS) | Suggestion: introduce pricing as a condition, check the EU and national perspective | 1 (P); 8 (SAS); 8 (PS) | 7 (IT); 7 (SAS); 8 (PS) | 6.5 |
| How could dynamic pricing schemes and smart charging be introduced in transport scenarios? | 6; 3 (PS); 9 (P) | Suggestion: | 6; 7 (PS); 5 (SAS) | 5; 5 (SAS); 5 (PS) | 5.66 |
| How can freight rail be improved to further increase its modal share to the detriment of road freight transport? | 7 (SAS); 7 (IT); 7 (P); 8 (PS) | Suggestion: | 7 (IT); 8 (SAS); 8 (PS) | 6 (IT); 8 (PS); 7 (SAS) | 7.3 |
| Urban area potential for mode shift (how to shape cities to stimulate modal change and use of soft modes)? | 7 (IT); 8 (PS); 7 (SAS); 8 pedro | Suggestion: sharing aspects in urban mobility | 6 (IT); 8 (SAS); 8 (PS) | 8 (PS); 8 (SAS); 6 (IT) | 7.4 |
| What is the role of systemic efficiency improvements in the overall transport as an enabler of energy efficiency? | 9 (SAS); 8 (PS); 8 (P) | Suggestion: | 8 (P); 9 (SAS) | 9 (PS); 8 (SAS) | 8.4 |
| What are important linkages between transport and other demand sectors in scenario building? | 2 (SAS); 5 (PS) | Suggestion: | 5 (SAS); 5 (PS) | 4 (SAS); 5 (PS) | 4.33 |

Questions (P) Rate and timing (incentives too) of transport decarbonisation- implications with infrastructure

Socio-economic aspects need to be accounted for. (ASA)

Figure 5. Co-designed and ranked questions for the industry sector, workshop 1.

#### 4.1.4. Scenario creation and adoption

Following the workshops, ECEMF modelling teams constructed scenario narratives based on the highest-ranked research questions from each workshop. The reasoning behind this approach is that the modellers are well informed about the limitations of their models, granting informed decisions on which research questions can be answered by their energy system and integrated assessment models. Several co-designed research questions had already been incorporated (in whole or in part) in scenarios developed by the ECEMF teams. Other research questions will be considered for analysis in future project steps.

Results from the scenario analysis conducted by the ECEMF modelling teams will be disseminated to policymakers and other stakeholders at national and international levels. This ensures that the scenario creation has adopted the main points from the co-designed research questions.

#### 4.1.5. Feedback loops and lessons learned

Feedback and lessons learned were key factors in driving the workshop programme development. First, the ECEMF project team provided feedback. It included feedback on the materials prepared in advance of each workshop, namely the presentations, agendas, and questions to stakeholders. Secondly, stakeholder feedback was gathered after each workshop through specific areas on the Miro boards where stakeholders were urged to provide feedback. Additionally, ECEMF project members were sent questionnaires on their experience with planning and executing the workshops. All views were then consolidated, and adjustments were made accordingly.

### 4.2. Outcomes of stakeholder workshops

The workshops resulted in a set of 82 research questions. Of these, 23 were designed during the first, 27 during the second and 32 during the third workshop. Exploring scenarios to answer all identified questions would require more resources and time than available, and some questions simply fall outside the scope of what the ECEMF models can represent. The three highest-ranked research questions for each thematic session are available in the supplementary materials.

# 5. Discussion

The literature review findings suggest that the typology of key stakeholder engagement concepts is fragmented within EU energy and climate research. This includes the definition of the term 'stakeholder', usually defined in various ways depending on context and field of research. The stakeholder engagement purpose is frequently described as co-creation, co-production, and co-design. While common concepts, they are rarely defined and often used interchangeably. In addition to the typology, the literature provides varying guidance on stakeholder engagement. This results from parallel developments of methodologies for stakeholder engagement and participation approaches. Identified commonalities among the approaches to stakeholder engagement include an explicit need for stakeholder identification, categorization, differentiation, and selection methods. These form the core of a wide range of frameworks [16, 42-44]. Overall, the literature on stakeholder engagement highlights a lack of consensus in the typology while following some common principles in terms of its implementation.

We performed a scoping review to answer the first research question. It included approximately 1500 open-access publications that were reviewed based on pre-defined exclusion criteria. Our findings show a lack of consensus on what constitutes a stakeholder within EU climate and energy research and that the term is rarely defined. These findings are not specific to energy and climate research. Instead, they support the broader stakeholder engagement literature presented in section 2.

Further review results highlight that engagement activities often lack a methodology for stakeholder identification and selection. Only two of the 11 selected articles implemented methodologies for stakeholder identification and selection, both applying the Prospex-CQI [55]. The lack of systematic approaches can adversely impact the quality and reproducibility of results. Without a documented and reproducible methodology, the approach becomes limited to its project implementation and could induce representation bias in the results.

Stakeholder categorization and selection methodologies based on knowledge and function [43], on interest-influence matrices [42, 44, 50], or through functional criteria and stakeholder roles combined with dimensions of interest and influence [23, 51], help reduce representation bias. Although negative impacts of stakeholder engagement are rarely brought up in the literature [18], some unintentional 'dark side' effects stem from misalignment, leading to conflict of interests. Considering the negative impacts, planners of stakeholder engagement workshops should aim to apply and improve available methodologies for stakeholder identification and selection to the extent possible.

Experiences from the ECEMF workshop programme development help us answer the second research question. The findings indicate that snowball sampling, when coupled with an online questionnaire, is an effective approach to obtaining diverse stakeholder representation. The workshop participants included attendees from 15 different countries with varying experience and fields of expertise. Overall, participants from academia, research institutions and intergovernmental organizations were represented to a higher extent compared to the private sector, non-profit organizations, and NGOs. This could be due to the nature of the topics discussed during the workshops. However, it may also be an artefact of the identification process, building a snowball sampling based on personal networks from within the ECEMF project. Nevertheless, the approach taken for identifying stakeholders led to enough stakeholders participating in each workshop to gain meaningful discussions and elicit stakeholder knowledge and feedback. The presented approach demonstrates the importance of preparing and accounting for the possibility of participant drop-offs. Adapting the workshop approach to fewer attendees is crucial since unforeseen hindrances can occur at short notice. If the drop-off is reported with a couple of days' notice, workshop planners can contact additional stakeholders to compensate for the drop-off.

The stakeholder selection process did not adhere to the methods proposed in the broader stakeholder analysis literature. This was unintentional and not planned for by the workshop organizers. We managed to proceed with all participants who signed up and accepted the invitations due to last-minute drop-offs and a sufficient capacity to host many stakeholders.

The stakeholder engagement process outlined in this article provided several important inputs to the ECEMF project. It offered a meeting point for researchers and policymakers to discuss the important topics relating to the EU energy transition. Contributions from private sector representatives and non-profit and non-governmental organisations further enriched the discussions. Out of these came a set of 82 co-designed and ranked research questions that make the foundation for the scenario narratives to be explored by the ECEMF modelling teams. The number of co-designed questions, the level of agreement on their policy relevance, and the ranking indicate that the approach taken was successfully executed.

As the approach was exploratory, many lessons were learned from the experience of planning and facilitating the workshops. Planning a workshop with two sessions on two separate days contributed to confusion among participants on when to attend. We recommend that special care be taken during the invitation if a similar approach is applied. Integrating feedback loops into the approach helped recognise shortcomings early in the process, allowing adjustments before the next workshop. An example includes the workshop format increase by 30 minutes, allowing more time for the breakout and plenary discussions. Further insights include that while younger generations are early adopters of new technologies, senior participants may not follow the same trend. This led to some participants being uncomfortable using Miro's features. It did not affect the output as the breakout sessions were facilitated. However, it increased the workload of the facilitators. This was adjusted through early identification, and the succeeding workshops included two facilitators in each breakout session.

The workshops included presentations by ECEMF members, which aimed to provide a deeper understanding of the project activities. While this is a practical approach to informing participants of the project activities, it can also introduce potential bias among the stakeholders.

Presenting results before identifying research needs risks introducing stakeholders to *the anchoring effect* [72]. The anchoring effect could have affected decisions about research needs, such as which research questions to focus on. Therefore, we suggest that facilitators and workshop planners carefully consider the contents presented to stakeholders before any ranking of priorities. Another potential impact is the *sleeper effect* [73], where familiarity with words or statements affects the stakeholders' decision-making. Repeatedly presenting unfamiliar but plausible statements leads to a greater inclination to accept those statements as true compared to statements not previously presented [74] as cited in [73]. To reduce these effects, we adjusted the approach in the third workshop, where we only presented a brief overview before the breakout sessions without showing sector-specific results. Given that only one workshop on the selected topics was conducted, the impact of this change was not measured.

While evaluating the co-designed questions we observed topics being highlighted across multiple thematic sessions and workshops. One example is the topic of hydrogen. Stakeholders raised questions regarding storage, utilisation, and sector coupling with hydrogen and its impact on emission reductions. This highlights the importance of hydrogen in current policy discourse among EU stakeholders. The EU goal for the 2020-2024 period is to add 6 GW [76] of electrolyser capacity for green hydrogen production. As of March 2023, the actual capacity is around 160 MW [77], indicating many challenges in the large-scale deployment. The phenomenon where stakeholders considered hydrogen a critical component of the transition, even though the current deployment of the technology is limited in the EU context, can be attributed to the judgmental heuristic called *availability* [75]. A better understanding of the biases caused by the *availability* heuristic could lead to better judgements in situations involving uncertainty. It should, therefore, be considered by stakeholder engagement facilitators to mitigate this type of bias.

The workshops generated 82 co-designed and ranked research questions, among which commonalties emerged, including a focus on trade-offs, the consequences of policies and incentives, EU production vs imports, and incorporating the climate change impacts. The co-designed questions presented in the supplementary materials should, therefore, be further investigated by the ECEMF project and other research initiatives capable of providing answers.

## 6. Conclusions

We conclude that the field of stakeholder engagement is fragmented in the literature but that some commonalities in engagement approaches exist. Different methodologies exist to support engagement planners in various engagement stages. This is particularly true for stakeholder identification and selection methods, such as the Prospex-CQI, snowball sampling, or influence-interest matrices. Despite an increase in the inclusion of stakeholder engagement for climate and energy research, few state-of-the-art articles suggest how to conduct stakeholder engagement within this field. With our scoping review, we identified that insufficient attention is given to the definition of the term 'stakeholder' and concepts such as 'co-creation', 'co-production', and 'co-design'. These terms are often used interchangeably, causing confusion as to what the concepts entail. We also identified that recent articles in the field of EU climate and energy research fail to define and implement critical stakeholder engagement methods such as stakeholder identification, categorization, and selection. Our study presents additions to the current literature by presenting a clear overview of the current research, as well as definitions for common stakeholder engagement concepts. The results suggest that stakeholder identification through a combination of snowball sampling and online questionnaires is an effective approach to reaching out to potential stakeholders. In the ECEMF case study, the workshop programme resulted in a wide range of stakeholders attending the various workshops, providing a meeting point for the co-design of 82 policy-relevant research questions. Stakeholder feedback indicated high agreement with the questions, and the active participation and engagement of the stakeholders confirmed that this approach had been successful. Additionally, we propose the presented workshop programme to be used by other researchers within the field, to be further improved and adjusted to different project needs. Lastly, we call for incorporating bias mitigation strategies in stakeholder engagement for EU climate and energy research, as climate change, the energy transition, and multi-stakeholder engagements include sources of uncertainty and bias.

## Funding

This project has received funding from the European Union's Horizon 2020 research and innovation programme under grant No 101022622. The funding source was not involved in the preparation of the article.

## CRediT author statement

**Emir Fejzić:** Conceptualization, Data curation, Formal Analysis, Investigation, Methodology, Visualization, Writing – original draft. **Will Usher:** Conceptualization, Funding Acquisition, Methodology, Supervision, Visualization, Writing – review & editing.

## Conflict of interest

The authors declare that the research was conducted without any commercial or financial relationships that could potentially create a conflict of interest.

## Acknowledgements

The authors would like to thank the stakeholders who participated in the workshops outlined in this paper for their contributions to the discussions, knowledge sharing through presentations, and the co-design process. Furthermore, we wish to express our gratitude to our project partners, who participated in preparing and facilitating these workshops.